\documentclass [12pt,preprint] {aastex}
\usepackage {natbib}
\usepackage {amsmath}
\usepackage {amstext}
\usepackage {latexsym}
\usepackage {xspace}
\begin{document}
%\draft
%\preprint{HEP/123-qed}
% --- /home/prasad/.lib/tex/vgarticle.sty ---
% =====================================================================
% Personal article style			v guruprasad,sep1999
% =====================================================================

% ---------------------------------------------------------------------
% General favourite definitions

% Orderly dates 
\newcommand\DateYMD{
	\renewcommand\today{\number\year.\number\month.\number\day}
	}

\newcommand\DateDMY{
	\renewcommand\today{\number\day.\number\month.\number\year}
	}

\newcommand\DateDmmmY{
	\renewcommand\today{
		\number\day\space
		\ifcase\month\or
		Jan\or Feb\or Mar\or Apr\or May\or Jun\or
		Jul\or Aug\or Sep\or Oct\or Nov\or Dec\fi
		\number\year
		}
	}

% Tables:
\newenvironment{mptbl}{\begin{center}}{\end{center}}
\newenvironment{minipagetbl}[1]
	{\begin{center}\begin{minipage}{#1}
		\renewcommand{\footnoterule}{} \begin{mptbl}}%
	{\vspace{-.1in} \end{mptbl} \end{minipage} \end{center}}

% Figures:
\newif\iffigavailable
	\def\figavailable{\figavailabletrue}
	\def\nofigavailable{\figavailablefalse}
	\figavailable% default

\newcommand{\Fig}[4][bh]{
	\begin {figure} [#1]
		\centering\leavevmode
		\iffigavailable\epsfbox {\figdir /#2.eps}\fi
		\caption {{#3}}
		\label {f:#4}
	\end {figure}
}

% Deflists:
\newlength{\defitemindent} \setlength{\defitemindent}{.25in}
\newcommand{\deflabel}[1]{\hspace{\defitemindent}\bf #1\hfill}
\newenvironment{deflist}[1]%
	{\begin{list}{}
		{\itemsep=10pt \parsep=5pt \topsep=0pt \parskip=10pt
		\settowidth{\labelwidth}{\hspace{\defitemindent}\bf #1}%
		\setlength{\leftmargin}{\labelwidth}%
		\addtolength{\leftmargin}{\labelsep}%
		\renewcommand{\makelabel}{\deflabel}}}%
	{\end{list}}%

% Equation numbering:
\makeatletter
	\newcommand{\numbereqbysec}{
		\@addtoreset{equation}{section}
		\def\theequation{\thesection.\arabic{equation}}
		}
\makeatother

% ---------------------------------------------------------------------
% General settings

\DateYMD			% the only rational default
\def\figdir{_figs}		% redefine this locally

% =====================================================================

% --- /home/prasad/.lib/tex/vgabbr.sty ---
% =====================================================================
% Personal abbreviations			v guruprasad,sep1999
% =====================================================================
% Scientific:

\def\arcdeg{\hbox{$^\circ$}}
\providecommand{\bold}[1]{\mathbf{#1}}
\newcommand\degree{$^\circ$}
\newcommand{\Qed}{$\bold{\Box}$}
\newcommand{\order}[1]{\times 10^{#1}}
\newcommand{\Label}[1]{\ \\ \textbf{#1}}
\newcommand{\Prob}[1]{\mathrm{\mathbf{Pr}}[#1]}
\newcommand{\Expect}[1]{\mathrm{\mathbf{E}}[#1]}

% Latinora:

\newcommand\ala{\emph{a la}\xspace}
\newcommand\vs{\emph{vs.}\xspace}
\newcommand\enroute{\emph{en route}\xspace}
\newcommand\insitu{\emph{in situ}\xspace}
\newcommand\viceversa{\emph{vice versa}\xspace}
\newcommand\terrafirma{\emph{terra firma}\xspace}
\newcommand\perse{\emph{per se}\xspace}
\newcommand\adhoc{\emph{ad hoc}\xspace}
\newcommand\defacto{\emph{de facto}\xspace}
\newcommand\apriori{\emph{a priori}\xspace}
\newcommand\Apriori{\emph{A priori}\xspace}
\newcommand\aposteriori{\emph{a posteriori}\xspace}
\newcommand\nonsequitor{\emph{non sequitor}\xspace}
\newcommand\visavis{\emph{vis-a-vis}\xspace}
\newcommand\primafacie{\emph{prima facie}\xspace}
\newcommand\ceterisparibus{\emph{ceteris paribus}\xspace}
\newcommand\adinfinitum{\emph{ad infinitum}\xspace}

% http://www.liv.ac.uk/education/hd/latin.html
\newcommand\circa{\emph{c.}\xspace}
\newcommand\ibid{\emph{ibid.}\xspace}		% previous citation
\newcommand\loccit{\emph{loc.\ cit.}\xspace}	% cited in the ref
\newcommand\opcit{\emph{op.\ cit.}\xspace}	% cited in the ref
\newcommand\viz{viz{}\xspace}			% videlicet - NO STOP

% http://www.tsolv.com/schools/lghs/clubs/latin/Latin_Abbreviations.html
\newcommand\ie{i.e.{}\xspace}			% id est
\newcommand\eg{e.g.{}\xspace}			% exempli gratia
\newcommand\etal{\emph{et al.}\xspace}		% et alii, et alibi
\newcommand\cf{cf.{}\xspace}			% confer (compare)
\newcommand\etc{etc.{}\xspace}			% et cetera

% General readability:
\newcommand{\figref}[1]{Fig.\ (\ref{f:#1})}
\newcommand{\figsref}[2]{Figs.\ (\ref{f:#1}-\ref{f:#2})}

\newcommand{\LD}{\begin{description}}
\newcommand{\DE}{\end{description}}
\newcommand{\LI}{\begin{itemize}}
\newcommand{\LE}{\end{itemize}}
\newcommand{\LN}{\begin{enumerate}}
\newcommand{\NE}{\end{enumerate}}
\newcommand{\VB}{\begin{verbatim}}
\newcommand{\VE}{\end{verbatim}\\}
\newcommand{\QB}{\begin {quotation}}
\newcommand{\QE}{\end {quotation}}

\newcommand{\Or}{\vee}
\newcommand{\Def}{\stackrel{\triangle}{=}}

% Documentation:
\def\ednote#1{\noindent== \emph{#1} ==}		% Editorial notes
\def\note#1{\noindent====\\\emph{#1}\\====}	% Yes notes!
\def\revising{\ednote{--------------- to be revised ------------------}}
\def\comment#1{}			% no comments!

% =====================================================================

% --- epsf.tex ---
% --- title.tex ---
\title {An optical solution of Olbers' paradox}
\author {V. Guruprasad}
\affil{IBM T J Watson Research Center, NY 10598, USA.}
\email{prasad@watson.ibm.com}

%\date{1999/1/7}
\begin{abstract}
% --- abs.tex ---
%
Shown is that contrary to common intuition, 
even an arbitrarily weak attenuating mechanism
is sufficient to make the background sky quite dark
independently of the size of the universe and the Hubble expansion.
Further shown is that
such an attenuation already exists in the wave nature of light
due to entrapment and diffusion from successive diffractions.
This is a fundamentally new mechanism to physics,
as illustrated by
application to
the solar neutrino attenuation,
galactic dark matter and
gamma ray bursts problems.
It not only provides a big bang-like cutoff,
but also appears to explain the appearance of
primeval, metal-deficient galaxies at high redshifts,
without deviating from the Olbers' premise of
an infinite universe.
\end{abstract}
\maketitle
% --- body.tex ---
%\clearpage
\section {Introduction}
\label {s:intro}

This historical paradox highlighted by Olbers almost two centuries ago
\citep{Olbers1826}
holds that
the night sky should have been as bright as the sun's disk
because
we should encounter rays from a stellar surface
in any direction we look.
More particularly,
the argument holds that
the stellar light must cover an increasing area $\propto r^2$
as its distance $r$ from its source increases,
its brightness  $J$ diminishing as $J_0/r^2$,
where $J_0$ is the stellar surface luminosity.
This is the well known inverse-square law for light,
and takes only the geometrical spread in 3-dimensional space
into account.
According to the paradox,
as we look at regions of the sky where the stars are very far,
they should each appear dimmer,
but a given solid angle would cover $\propto r^2$ stars,
assuming a uniformly populated infinite universe.
The luminosity of the sky should therefore be
$\propto (J_0 / r^2) \times r^2 = J_0$,
meaning that
the background sky should be so bright
that the stars should be indistinguishable against it.
I present two fundamental results,
first,
that an infinitesimally small attenuation $\sigma > 0$,
barely enough to change the propagation law to the form
$J_0 \, e^{- \sigma r}/r^2$,
suffices to solve the paradox,
and second,
that such an attenuation happens to be inherent
in the wave nature of light.

The results are unintuitive because
almost any kind of attenuation leads to this form, and
absorption and scattering by dust
have been considered inadequate in the past.
Harrison argued
\citep{Harrison1965}
that the dust would eventually attain thermal equilibrium with
the stars and effectively stop absorbing more energy.
Since the radiation too would eventually reach equilibrium,
scattering of itself cannot solve the paradox either.
The standard model offers three plausible solutions
%\begin{list}{-}{\topsep 0pt \itemsep 0pt \parsep 0pt}
%\item
	that the Hubble flow causes the light to lose energy
	well in excess of the inverse-square attenuation,
%\item
	that the universe is too young for thermal equilibrium,
and
%\item
	that the universe is as such finite.
%\end{list}
Wesson has shown that
the first would actually contribute less
\citep{Wesson1987},
so finiteness of the universe,
in both age and extent,
is currently believed to be the reason for
the darkness of the night sky.
I shall show that
a similar cutoff occurs because of the inherent attenuation,
that it necessarily leads to
a further mechanism of spectral modification
that would make the most distant galaxies appear primeval,
which is again known and currently attributed to the big bang.
In both cases,
I exploit the \emph{smallness} of the mechanisms
needed for the respective effects.
The approach is only possible
because of the exponential factor in the attenuated propagation law,
but more importantly,
it demonstrates that
the small orders routinely left out in the approximations
applicable on earth could be significant physics
on the cosmological scale.
To further emphasise
the error in letting approximations dictate our reasoning,
I shall show that
an even smaller order of attenuation due to the same mechanism
would explain the missing solar neutrino flux,
currently attributed to neutrino oscillation,
and could mean new insight of a fundamental kind
in the physics of weak interactions.

Unlike the dust effects,
the attenuation of present concern
depends only on the presence and not
the thermal state or other properties of matter, and
is therefore immune to thermal equilibrium.
It results from
a diffuse entrapment of radiation due to
successive diffraction and gravitational deflections
that continually turn a portion of the wavefront.
While the occurrence of successive deflections is known,
for example, in Laue diffraction theory,
its implications to astrophysics 
have not been examined in previous treatments,
such as in the context of extinction by dust
\citep[p149-153]{Spitzer1978},
possibly because
one ordinarily thinks of diffraction
as carrying wave energy around obstructions,
increasing rather than diminishing the net power flow. 
This would be the case in a hypothetical universe
where the sources are assumed to be
behind a plane of diffracting obstructions,
but the notion does not really extend to
the three-dimensional universe involved in the paradox,
where the stars themselves
are the obvious obstructions
to each other's light.
The enhancing property turns out to be intransitive
because the successive deflections can then keep
a fraction of the radiation from ever reaching
its original destination.

Another reason why this result was unobvious is that
Fraunh\"ofer's approximation is invariably assumed
because of the immense distances involved.
A solution to Olbers' paradox then appears to be ruled out because
the loss of the direct rays from a star
due to \enroute diffraction by an angle say $\theta$,
would be made up by the rays from another star
behind the diffractor at an angle $-\theta$.
While the argument is somewhat weaker than enhancement,
it reveals the error in our past intuition,
because the probability $p(r_s)$ of finding a compensating star
\emph{at the same or less distance $r_s$} behind the diffractor
depends on $r_s^2/r^2$,
% where $r_s$ denotes the distance from the diffractor to the source,
which is certainly less than unity.
The approximation assumes that 
both the sources and the observer are infinitely far
from the diffracting object,
\ie
$r_s \rightarrow \infty \text{ and } r \rightarrow \infty$,
in which case the ratio does not matter.
These conditions are, however, applicable only
in the vicinity of a given diffractor,
and cannot be legitimately applied
when the obstructing objects are distributed over
the same scale of distances as the sources.
In the presence of an attenuation $\sigma$,
the distance $r_s$ matters because
the compensating source becomes more likely to be farther
by the triangle theorem,
and therefore likely to be dimmer.

In both the real universe and Olbers' scenario, therefore,
the light from a distance source
located on a geometrically unobstructed straight line from us
does get diminished by diffraction
due to obstructions lying off the straight line.
This loss would be compensated,
as in the Fraunh\"ofer case,
if enough diffracted light from elsewhere
could rejoin the straight line path.
Traditional wisdom suggests that
the repeated deflections be treated as
as a random walk leading to a slow diffusion of the photons,
which would not yield a net reduction of the average luminosity.
However,
there are problems with this view,
because while it is reasonable to assume that
the diffracting objects are randomly distributed,
their gross motions are not random and
are quite slow in relation to the interstellar distances.

What we have is an essentially static pattern
lacking the temporal randomisation of direction
needed to qualify it as random walk.
More particularly,
all efforts to simulate thermalisation with fixed dynamical models
have consistently led to persistent oscillatory states,
since the very first attempt in 1953
by Fermi, Pasta and Ulam
\citep{FPU1955,Fillipov1998},
showing that mere complexity of dynamical structure
is not enough for assuming diffusion.
Moreover,
any static pattern necessarily contains circulations,
which might not only explain the FPU problem,
but in our case,
trap some of the light virtually forever.
Harrison's argument cannot be applied to such states
because the circulations would be centrally dependent on
the individual sources, and
each of which presumably has a finite lifetime
even in Olbers' scenario.
We do expect most of the ``trapped'' energy to eventually diffuse out,
but we have no basis to assume that \emph{all} of it will.
Rather,
we can expect a portion of the energy to get absorbed
or turn into matter,
given that the attenuated propagation law is
already characteristic of the Klein-Gordon equation
$(\nabla^2 + \sigma^2 \partial^2 / \partial t^2) \, \psi = 0$.
The latter is simply the quantum version of
the relativistic argument that radiation
when retarded to effective speeds less than $c$
should exhibit rest mass, and
relates to our treatment of the solar neutrino problem.
We can thus be certain of a net attenuation $\sigma > 0$, and
this suffices, as shown next, to solve the paradox.

\section {Solution of the paradox}
\label {s:olbers}

As stated,
our key argument is that any attenuation whatsoever,
so long as it operates on the large scale,
solves the paradox independently of the standard model.
We seek an attenuation $\sigma$ (dB m$^{-1}$),
such that the propagation law for light changes
from $r^{-2}$ to $r^{-2} \, e^{-\sigma r}$,
which reduces the background brightness of the sky to
\begin{equation} \label{e:isigma}
	J_\sigma
	= \int_{\infty}^0 J_0 \, e^{-\sigma r} \, dr
	=
		J_0 / \sigma .
\end{equation}
While the integral is well known,
the solution is not immediately obvious,
largely because the value superficially resembles $J_0$,
%the unattenuated result in the paradox,
making it look as if we need a very large $\sigma$,
of the order of at least $130$~dB $\equiv 10^{13}$
\citep[p.24-25]{RoachGordon}
to get a dark background sky,
and hitherto seemed impossible without the big bang theory.

The resemblence is misleading
because the unattenuated $J$ differs in dimensions from $J_0 / \sigma$,
as the latter has the dimensions of \emph{luminosity $\times$ distance}.
For legitimate comparison,
we must express $J$ in exactly same dimensions,
hence in the statement of the paradox,
the ``Olbers luminosity'' $J$,
which one may informally think of as the brightness of the sun's disk,
corresponds not to $\sigma = 1$ but to $\sigma = 0$,
\ie
\begin{equation} \label{e:iolbers}
	J \equiv
	\lim_{\sigma \rightarrow 0}
	\int_{\infty}^0 J_0 \, e^{-\sigma r} \, dr
	=
	\lim_{\sigma \rightarrow 0}
		J_0 / \sigma ,
\end{equation}
which is infinity.
Conversely,
when we calibrate with respect to the sun's disk,
the attenuated background sky should be infinitesimally dim,
for any $\sigma > 0$.
To appreciate why this should be so,
consider how bright the background needs to be
in order to match the sun's disk.
In Olbers' argument,
whichever direction we look in,
our line of sight must meet a stellar surface and
at any finite angular resolution $\theta$,
the brightness should correspond to
the number of stars included within the solid angle $\theta$.
Accordingly,
the observed brightness
would be $J_\sigma \equiv J_0 \, e^{-\sigma r}$
along that direction, and
\emph{%
that set of stars then needs to be $e^{\sigma r}$ times brighter
than the sun in order to match $J_0$}.
Clearly,
it does not matter if $\sigma$ is very small;
as long as it is nonzero,
we have an effective cutoff of the observable universe at
\begin{equation} \label{e:icutoff}
	r_n = \sigma^{-1} \log (J_0 / J_n)
\end{equation}
for a finite $\sigma$,
where $J_n$ represents the background noise in the measuring process.
By eq.\ (\ref{e:icutoff}),
the standard model cutoff,
corresponding of an age of the universe of $\sim 15$~Gy,
is equivalent to
\begin{equation} \label{e:rcutoff}
	\sigma \approx 130 \text{ dB/} 15 \text{ Gy}
	= 9 \times 10^{-25} \text{~dB m$^{-1}$}
	.
\end{equation}
This is far less than what one might naively expect
from the form of $J_\sigma$ (eq.\ \ref{e:isigma}),
and amounts to a mere
$8 \times 10^{-9}$~dB per light-year.
As remarked in the introduction,
it is the \emph{smallness} of the attenuation
needed to explain the apparent big bang cutoff
that makes it impossible to rule it out
on the basis of terrestrial physics.
Furthermore,
eq.\ (\ref{e:rcutoff}) is as yet an upper bound,
because we ignored the diffraction from nonluminous bodies 
as well as the gravitation of these and the visible objects,
which contribute substantially to the deflections.
We also ignored the dust extinction in our own neighbourhood
\citep[ch.4]{RoachGordon},
to which Harrison's argument again does not apply,
and the impact of the Hubble flow,
which together account for a good part of the $130$~dB.
Note that Harrison's argument remains valid
for dust on the large scale,
and the subtlety that overcomes it for
the diffractive scattering,
described in the next section,
is the coherence and dependence of
the diffuse circulatory states on their respective sources.

%\clearpage
\section {Scattering approximation}
\label {s:scat}

Eq.\ (\ref{e:iolbers}) establishes our principal point
that only an extremely small attenuation is needed
to reproduce the big bang cutoff.
It remains to be shown that
such an attenuation is indeed possible and likely
from successive diffraction.
As explained in \S\ref{s:intro},
we are concerned with multiple interstellar hops
that are each much larger than stellar diameters,
so that the Fraunh\"ofer theory can be applied to
the individual diffractions.
We may more particularly treat the stars as point objects,
representing the diffraction by
an angular spreading function $f(\phi, \delta)$,
$f \ge 0 \; \forall \; \{ \phi, \theta \}$,
applicable to a parallel beam of incident light,
giving
\begin {equation} \label {e:spread}
	J (\phi, \delta)
	=
	J_0 f (\phi, \delta) 
\text { and }
	\int_{\phi = 0}^{2\pi}
	\int_{\delta = 0}^{\pi}
		f (\phi, \delta)
	\, d\phi
	\, d\delta
	= 1 ,
\end {equation} 
for the diffracted light,
$\phi$ being the azimuthal angle around the beam axis and
$\delta$, the angular spread from the axis.
We shall now examine the special treatment
necessary to account for successive diffractions.

The spread function $f$ is rather like
the differential scattering function $\sigma (\Omega)$
of the Rutherford model
\citep[\S3-20]{Goldstein},
but several important differences must be noted.
Firstly,
we are concerned with continuous wavefunctions, not particles,
so that the ``scattering'' itself is not at all probabilistic;
the only probability inherent in the model
is the stellar distribution.
Secondly,
the scattering function $f$ depends on the $\lambda/D$ ratio,
where $\lambda$ is of course the wavelength and
$D$, the mean stellar diameter.
This is treated in more detail in terms of
Fresnel-Kirchhoff theory in Appendix \ref{a:point}.
More importantly,
the result of interest is the net attenuation and
not a cross-section of interaction.
It is common to use $\sigma$ for both,
but the second notion is not of concern here.
We shall now formalise the cumulative forward loss
and the nonzero ``backscatter'' occurring at each encounter.

The second of eqs.\ (\ref{e:spread}) represents
the conservation of energy at each diffractive encounter,
since it means that
\begin {equation} \label {e:paraxial}
	\int_{\phi = 0}^{2\pi}
	\int_{\delta = 0}^{\pi}
		J (\phi, \theta)
	\, d\phi
	\, d\delta
	=
	J_0
	.
\end {equation} 
We start by applying $f$ to
a bundle of rays incident on a first star.
Since the rays spread as if from a point source,
they acquire a $1/4 \pi r^2$ spreading loss,
even if we had started with a parallel bundle,
which would be equivalent to a planar wavefront.
A second encounter after a distance $r_1$ therefore yields
\begin {equation} 
\begin {split}
	J (\phi, \theta)
	&=
		\frac{J_0}{4 \pi r_1^2}
	\int_{\delta = 0}^{\theta}
		f (\phi, \theta - \delta)
		f (\phi, \delta)
	\,
		d\delta
	\\
	&\equiv
	\frac{J_0}{r_1^2}
		\,
		g_1 (\phi, \theta)
\end {split}
\end {equation}
where $g_1$ denotes the first cumulative integral 
\begin {equation}
	g_1 (\phi, \theta) =
		\frac{1}{4 \pi}
	\int_{\delta_1 = 0}^{\theta}
		f (\phi, \theta - \delta_1)
		f (\phi, \delta_1)
		\, d\delta_1
	.
\end {equation} 
We thus obtain the recursive set of integrals
\begin {equation} \label {e:recursive}
\begin {split}
	g_0 &\equiv f
\quad
	\text { and }
\\
	g_n (\phi, \theta)
	&=
		\frac{1}{4 \pi}
	\int_0^{\theta}
		f (\phi, \theta - \delta)
		\,
		g_{n-1} (\phi, \delta)
		\, d\delta
\;
	\text { for }
		n = 1, 2, ...
\end {split}
\end {equation} 
describing $n$ successive encounters as
\begin {equation} \label {e:succ}
%\begin {split}
	J (\phi, \theta)
	=
	J_0 \, g_n (\phi, \delta)
	\,
	\prod_{j = 1}^{n}
		\frac{1}{r_j^2}
%\\
	\approx
	J_0 \,
		g_n (\phi, \delta)
	\,
	\bar{r}^{-2n}
%\end {split}
\end {equation} 
where $r_j$ are intervening distances, and
$\bar{r}$, the mean distance between stars.
Integrating eq.\ (\ref{e:succ}) over $\phi$, and
summing over the contributions from one or more encounters,
we obtain the total ``gain'' at $\theta$ from the initial beam as
\begin {equation} \label {e:cumsucc}
\begin {split}
	g (\theta) 
	&=
		\frac{g_0 (\theta)}{\bar{r}^2}
	+
		\frac{n_1 g_1 (\theta)}{\bar{r}^4}
	+
		\frac{n_2 g_2 (\theta)}{\bar{r}^6}
	%+
	%	\frac{n_3 g_3 (\theta)}{\bar{r}^8}
	+ ...
	\\
\text { where }
	g_n (\theta)
	&\equiv
		\int_0^{2\pi} g_n (\phi, \theta) \, d\phi
	.
\end {split}
\end {equation} 
Here, $n_j > 1$ 
denote the mean number of parallel paths
corresponding to the range of $\phi$ for $j$ encounters, and
$g_j (\theta)$ are the per-path contributions.
It should be at least intuitively clear that
$n_j$ would be an increasing combinatorial function of $j$,
offsetting the increasing geometrical attenuation 
from the denominator ($\bar{r}^{-2j}$).

Eq.\ (\ref{e:cumsucc}) reveals an interesting property
of the successive diffractions:
that the $g_j (\theta)$ increase with $j$
for precisely the reason that
non-paraxial angles are ordinarily ignored
in terrestrial optics --
at large $j$'s,
small incremental angles $\delta$
become significant in the integrand,
so that
\begin {equation} \label {e:limgj}
	\lim_{j \rightarrow \infty}
		g_j (\theta)
	\equiv
	\lim_{\delta \rightarrow 0}
		\left[
			\frac{1}{2\pi}
			\int
				f (\phi, \delta)
			\,
			d\phi
		\right]
			^{\theta / \delta}
	\approx
		1
\end {equation}
as the integral converges to the zero-th order beam. 
Eq.\ (\ref{e:limgj}) represents the observation that
a succession of small diffraction angles
adds up to a large total deflection with significant amplitude,
since the component deflections are paraxial.
Three encounters each of $\pi / 3$,
for example,
suffice to make a back contribution, and
the direct backscatter $g_0 (\theta \approx \pi)$
suffers no $1/\bar{r}^2$ attenuation
as it involves no intermediate hops at all.
As a result,
nonzero total reflection is generally guaranteed and
the increasing $n_j$ also partly compensate for
the $1/\bar{r}^{2j}$ loss.
$g_0 (\pi)$ is generally ignored in terrestrial optics
because the $\cos^2 (\theta)$ factor is more pronounced
over small distances;
this too is inappropriate over the cosmological scale of distances.
As stated in \S\ref{s:intro},
motions of the surface of the diffracting star,
although many orders larger than the wavelength of light,
is not significant in the present context,
principally because
$f$ represents angular distribution of the radiant power
and is unaffected by phase fluctuations.

The presence of a net backscatter $g(\pi)$ means that
light from sources in the observer's own neighbourhood
will tend to lighten the sky.
This is an expected effect of scattering due to dust,
for example:
Roach and Gordon estimate that
for an observer near the centre of our galaxy,
the night sky should be relatively opaque
\citep{RoachGordon}.
This does not contradict
either the paradox or Harrison's argument,
but the diffractive backscatter potentially does
because it would be spread over cosmological distances.
Our result remains intact
because the backscattered light would itself be again subject to
repeated diffractions and suffer the same attenuation.
The totality of ``local'' sources
we need to consider would cumulatively increase as $r^3$,
but gets overcome by
the exponential attenuation $e^{-\sigma r}$ within a limited $r$.

%\clearpage
\section{Applications}
\label{s:app}

As mentioned in \S\ref{s:intro},
our mechanism has several interesting applications,
beginning with the manifestation of rest mass of the photons
because of the Klein-Gordon form of the attenuated propagation law.
This suggests that
our trapped circulatory photon states 
could be responsible for a portion of the dark matter
especially if the universe were very old.
A related observation is that
the density of circulatory states would be proportional to
the density of particulate matter,
because the deflections would become available at smaller distances
in dense neighbourhoods.
We would expect the attenuation to be greater
within galaxies than in the sparser regions between them,
and the resulting ``optical dark matter'' distribution
to be consistent with
the overall galactic dark matter
indicated by their rotation profiles,
as the circulatory states would presumably add up
with radial distance from the galactic centre.
At present,
we do not know how to estimate
this optical component of dark matter.

The attenuation might also explain
the intensity of gamma ray bursts from galactic regions.
For a given aperture between the stars and other objects
surrounding the line of sight,
the basic principles of diffraction dictate that
gamma rays would suffer
far less diffractive loss than visible light
because of their shorter wavelengths.
The observed gamma ray burst intensities
should therefore be much more than
we might expect from the visible brightness.
While the reasoning appears to be in the right direction,
it remains to be substantiated.

Another high energy scenario consistent with our theory is
the known attenuation of the solar neutrino flux,
which, as mentioned in \S\ref{s:intro},
is currently attributed to quantum oscillation.
Given that the mass of the sun is
$M_\odot \approx 2 \times 10^{30}$~kg and
the average mass of a nucleon,
$1.67 \times 10^{-27}$~kg,
we estimate
there are $1.2 \times 10^{57}$ nucleons within the sun's volume
$\approx 1.4 \times 10^{27}$~m$^3$,
which allows roughly $1.2 \times 10^{-30}$~m$^3$ per nucleon.
This means that
a neutrino would encounter a nucleon
every $1.06 \times 10^{-10}$~m,
or up to $6.6 \times 10^{18}$ times
from the centre to the surface.
This mean free path between nucleons
is the deBroglie wavelength for a particle of about $12$~keV,
so some diffraction appears to be inevitable. 
The observed attenuation, about two-thirds,
must result from $O(10^{18})$ compoundings,
assuming, as an approximation,
that all neutrinos originate from near the centre;
this yields
$\sigma \approx \exp [{10^{-18} \log (2/3)}]
\approx \exp (4 \times 10^{-19})$
or $10^{-19}$~dB per encounter,
as the diffractive attenuation needed
to explain the observed loss.
Once again,
it is the \emph{smallness} of the necessary attenuation
that makes it plausible as the cause, and
we speculate that
this could be the wave-theoretic reason for
the breaking of symmetry that leads to weak interactions.

Our last application is to
the appearance of primeval galaxies
at high redshift factors $z$, and
comes from observing that
the selective absorption of certain frequencies by matter
close to the line of sight should trigger 
diffractive loss $\sigma' \approx \sigma$
at the absorbed frequencies.
This should result in a subtractive spectral modification of
the light from more distant matter, and
the spectral subtraction by our own galactic neighbourhood,
which is well evolved and metal-rich,
should make more distant galaxies appear
metal-deficient and primeval.
Only an extremely small $\sigma'$ is needed, once again, to explain
the almost total metal-deficiency of the primeval galaxies.
As $\sigma'$ must result from
selective absorption by obstructions
that must be transparent at other frequencies,
\ie from a different set of obstructions, 
$\sigma'$ cannot be identical to $\sigma$.
But it would be of the same order,
leading to a similar cutoff $r_n' \approx r_n$
(eq.\ \ref{e:icutoff}),
because
denser regions of dust that invariably surround opaque bodies
are likely causes of selective absorption.
Importantly, we may argue that
the most distant galaxies can appear mature
\emph{only if there were a uniform mix of young and old galaxies
all the way right up to our own galaxy},
but this would contradict the standard model.
This does not mean unexpected support for the big bang theory,
however,
because we could merely be in
a relatively small region of the Olbers' universe,
of radius at least $r_n$,
that formed at or before
the big bang indicated by the Hubble redshift.

\acknowledgements

I thank Shai Ronen, Gyan Bhanot, A Joseph Hoane and
most of all, Bruce Elmegreen,
for valuable discussions in the context.

% end
\appendix
% --- xpoint.tex ---
\section {Point diffraction}
\label{a:point}

The usual treatment for diffraction due to a small region
concerns apertures in an opaque screen,
but our stars are in effect
obstructions in an otherwise empty space.
If $U_s$ represents the diffracted amplitude due to a star, and
$U_a$, that due to an aperture of the same size,
we have by Babinet's principle that
	$U_s = U_0 - U_a$,
where
$U_0$ represents the unobstructed incident wavefront.
In holography, for instance,
we can use this theorem to compute
the resulting interference between the direct rays ($U_0$)
and the diffracted rays ($U_s$) from a point obstruction.
In the present context, however,
we are not interested in the local interference patterns,
but only in the direction of the power flow,
so the interference between
direct and diffracted wavefronts is irrelevant.
We do sum over rays going in the same direction $\theta$
in eq.\ (\ref{e:cumsucc}),
but $\bar{r} \gg \lambda$,
and in any case,
there is enormous thermal jitter from our stellar obstructions,
which would randomise the phase between the summed paths,
hence interference effects can again be ignored,
even if our initial rays happened to be quite coherent. 

We do need to consider phase
within narrow bundles of rays, however, 
in order to obtain the wavelength dependence of the diffraction.
Under the far-field assumption $\bar{r} \gg D$,
$D$ being the mean diameter of our stellar obstructions,
the incident wavefronts before diffraction would be
not only almost planar,
but likely to be temporally coherent
across a sizeable fraction of the stellar surface.
We shall limit ourselves to pure Fourier components,
avoiding questions of incoherence,
because issues of decoherence and wavepacket dispersal
can be considered in terms of these components.
With these assumptions,
we may apply Kirchhoff's condition
that the conditions in a stellar diffracting region $\mathcal{A}$
will not be much affected by the surrounding matter,
which yields the usual Fresnel-Kirchhoff diffraction formula
\cite[\S8.3.2]{BornWolf}
\begin{equation} \label{e:FresnelK}
	U (\theta) =
	- \, \frac{i}{2 \lambda}
	\int_{\mathcal{A}}
		[ \cos (\hat{r}) - \cos (\hat{s}) ]
		\frac{ e^{i k (r + s)} }{rs}
		\, dS
	,
\end{equation} 
where $s$ and $r$ are distances
to the source and to the point of observation, respectively.
We cannot, however, afford to replace the cosines
with a single $\cos (\theta)$ factor
as in the traditional theory,
%\cite [\S8.3.3] {BornWolf},
since this would be proper only for paraxial rays.
To correctly compute $U$ at large diffraction angles $\delta$,
we need to retain the two-cosine form
in our Fraunh\"ofer approximation, obtaining 
\begin{equation} \label{e:Approx}
\begin{split}
	U (\theta)
	&=
		- \, \frac{i \, [ \cos (\hat{r}) - \cos (\hat{s}) ] }
			{2 \lambda rs}
		\int_{\mathcal{A}} e^{i k (r + s)} \, dS
	\\
	&\equiv
		- \, \frac{i \omega \,
			[ \cos (\hat{r}) - \cos (\hat{s}) ] }
			{2 \pi \, c}
			\frac{ e^{i k (r + s)} }{rs}
		\int_{\mathcal{A}}
			e^{i k (\xi + \eta)} \, dS
	,
\end{split}
\end{equation} 
$\xi$ and $\eta$ being the direction cosines as in prior theory.
A remaining problem is, of course,
that the above formulae are meant for
diffraction from an aperture, and
not an opaque disk presented by each star,
but Babinet's principle assures us that
the resulting $f$ will be identical.
Assuming our stars to be generally circular,
we find that
the integrand would have the form $2 J_1 (ka\theta) / ka\theta$,
where $k \equiv 2\pi/\lambda$ and $a$ is the radius of the aperture,
yielding
\begin{equation} \label{e:circ}
	f (\theta)
	\sim
	| U (\theta) \bar{r} |^2
	=
		\frac{ \omega^2 [\cos (\hat{r}) - \cos (\hat{s})]^2 }
			{ ( 2 \pi c )^2 }
		\,
		\left[
			\frac{ 2 J_1 (k a \theta) }{ k a \theta }
		\right]^2
	,
\end{equation} 
since the $(rs)^2$ factor is separately accounted for
by $\bar{r}$ in \S\ref{s:scat}.
 
As a result,
$f(\theta)$ is non-zero \emph{almost everywhere},
over the entire range of $\theta$, which includes $\pi$.
This ``backscatter'' is a reflection from space,
due to the local perturbation in the impedance of space
caused by the off-axis diffracting star, and
is definitely nonzero for any nonzero solid angle
around $\theta = \pi$.
The angular spread is more strongly governed
by the phase factor than by $\omega$,
so that
a given gap between nearer obstructions will be more open
to gamma rays than to visible light,
as remarked in \S\ref{s:intro}.

% === cat *.bbl ===

% --- figs.tex ---
\end {document}